\documentclass{jaa}
\usepackage{natbib}
\usepackage[utf8]{inputenc}
\DeclareUnicodeCharacter{2212}{-}
\usepackage{graphicx}
\usepackage{pdflscape}
\usepackage{amsmath,siunitx}
\usepackage{hyperref}

\begin{document}\sloppy

\title{Surface activity of rapidly rotating stars from simultaneous X-ray and UV observations with \emph{AstroSat}}

\author{Lalitha Sairam\textsuperscript{1,*}, Utkarsh Pathak\textsuperscript{2,3} and Kulinder Pal Singh\textsuperscript{3,4}}
\affilOne{\textsuperscript{1}School of Physics \& Astronomy, University of Birmingham, Edgbaston, Birmingham B15 2TT, UK.\\}
\affilTwo{\textsuperscript{2}Indian Institute of Technology Bombay, Powai, Mumbai-400076, India.\\}
\affilThree{\textsuperscript{3}Department of Physical Sciences, Indian Institute of Science Education and Research Mohali, Sector 81, SAS Nagar, Manauli PO 140306, India.\\}
\affilFour{\textsuperscript{4}Tata Institute of Fundamental Research, Homi Bhabha Road, Mumbai 400005, India.}


\twocolumn[{

\maketitle

\corres{L.Sairam@bham.ac.uk}

\msinfo{July 2023}{}

\begin{abstract}

Our study focuses on analysing the coronal, transition and chromospheric activity of four rapidly rotating stars located within 50 pc in the solar neighbourhood. We have used the multi-wavelength capabilities of \emph{AstroSat}, to investigate the outer atmospheres of AB~Dor, BO~Mic, DG~CVn and GJ~3331. These stars, classified as M and K type active stars, are known for their short rotation periods, leading to increased surface magnetic activity. Our soft X-ray observations provide the coronal properties such as emission measures, temperatures and elemental coronal abundances.  We report the detection of X-ray flares from AB~Dor, BO~Mic, and DG~CVn, while UV light curves reveal flares in both BO~Mic and DG~CVn.   
\end{abstract}


}]


\doinum{12.3456/s78910-011-012-3}
\artcitid{\#\#\#\#}
\volnum{000}
\year{0000}
\pgrange{1--}
\setcounter{page}{1}
\lp{1}

\section{Introduction}
\label{sec:intro}

Low-mass stars possess stratified atmospheres consisting of the photosphere, chromosphere, transition region and corona. Each of these atmospheric layers has a characteristic temperature and features associated with magnetic activity \citep{Kochukhov_2021, Hindman_2022}. In general, the Sun is considered to be a prototype of a low mass main-sequence star, and we often extrapolate the solar atmospheric knowledge to other stars. The first evidence of solar magnetic activity was provided by the presence of photospheric spots also known as sunspots (\citealt{clark_1978}, \citealt{Wittmann_1987}, \citealt{usoskin_2017}, and references therein). The observations of sunspots were later complemented by the observations in multiple wavelength bands such as ultraviolet, X-ray and radio wavebands tracing different solar atmospheric layers and their associated activity phenomena \citep{Hathaway_2015}. Additionally the different layers of atmosphere may not be physically disconnected, for instance, during a flare the material from the chromospheric layer is transported to the corona causing the coronal density and metallicity to change temporarily (\citealt{sylwester_1984}). However, several studies of low-mass stars have shown that such spatial correlations between the different layers of the atmosphere and their associated magnetic activity phenomenon may or may not be similar to that of the Sun. The heating mechanism of the outer layers of the atmosphere both in the stars and the Sun still remains a puzzle \citep{Toriumi_2022}.

\begin{table*}
\centering
\caption{Properties of stars observed with \emph{AstroSat}}
\label{tab:tab1}
\begin{tabular}{llllllc} 
\hline\hline
 &  AB~Dor & BO~Mic & DG~CVn & GJ~3331  \\ [0.5ex] 
\hline
Spectral type & K0V & K3V   & M4Ve & M1.5  \\
V (mag)       & 7.0  & 9.34 & 12.02 & 10.41   \\
Distance* (pc) & 14.85 $\pm$ 0.10 & 51.02 $\pm$ 1.50 & 18.29 $\pm$ 0.11 & 19.83 $\pm$ 0.01 \\
P$_{\mathrm{rot}}$ (d) & 0.514 & 0.380 & 0.2683 & 0.34/9.8  \\
$v \sin i$ (km s$^{-1}$) &91$\pm$1 & 
134$\pm$10 & 55.5 & 5.2\\
R$_{\star}$ [R$_{\odot}$] & 0.96$\pm$0.06&1.06$\pm$0.04& 0.46&0.80$\pm$0.05\\
\hline
\end{tabular}
\begin{tablenotes}

Note: Distances are based on parallax measurements in Gaia EDR3 \citep{gaia_collab,gaia_EDR3}. DG CVn distance is calculated from parallax measurement in Gaia DR2 \citep{gaia_collab, gaia_DR2}. Radius are based on \cite{drake_2015}, \cite{dunstone_2006}, \cite{osten_2016}, and \cite{messina_2014}.
\end{tablenotes}

\end{table*}

The hot coronal X-ray emission of low mass stars arises from the magnetically confined plasma with temperatures higher than 10$^6$ K (\citealt{guedel_2004}). Systematic studies of low mass stars in X-rays shows that many of these stars are considerably more active than the Sun and also far more rapidly rotating (\citealt{pallavicini_1981}). Stars with short rotation periods exhibit activity-related features such as spots, flares, enormous emission in activity sensitive chromospheric lines (Ca II and H$\alpha$), X-rays and ultraviolet wavelengths. In comparison to the Sun with $\mathrm{\frac{L_{X,Sun}}{L_{bol, Sun}}\sim10^{-6}}$ rapid rotators are known to possess a so called saturated corona with the $\mathrm{\frac{L_X}{L_{bol}}\sim10^{-3}}$ \citep{garcia_2008, lalitha_2017}. This raises a concern in what respect do the rapid rotators differ from the Sun to accommodate the excess X-ray emission from their coronae.

Our current knowledge of the coronae and outer atmospheres of rapid rotators are mostly based on individual measurements in either X-rays or UV. The UV emission is mostly observed in the transition region whereas X-ray emissions are associated with the coronal region, leading to an invariable connection between the two regions during enhanced activities. While the near-ultraviolet (NUV) band is dominated by the chromospheric emission line the far-ultraviolet (FUV) comprises  emission lines from the transition region. 
\cite{stelzer_2013} investigate the dependence of the UV excess on the X-ray flux using existing archival data. They found a strong correlation between coronal activity and emission in the transition region. 
To unravel this connection, we are carrying out a systematic survey of fast rotators simultaneously with the Ultra Violet Imaging Telescope and the Soft X-ray Telescope instruments onboard \emph{AstroSat}. 
We used \cite{cutispoto_2002} to identify fast rotators in the solar-neighbourhood. In this study we focus on four stars: AB~Dor, BO~Mic, DG~CVn, and GJ~3331. These stars share common traits of high magnetic activity, including high rotational velocities, and enhanced chromospheric and coronal emissions. The primary difference lies in their spectral classification, with AB Dor and BO Mic classified as K dwarfs, while DG~CVn and GJ~3331 are classified as M dwarfs. We provide a detailed description of each star and their known properties are given below. The fundamental stellar properties of all our targets are given in Table \ref{tab:tab1}.

\begin{enumerate}
    \item AB Dor: A young active K dwarf star ($\sim$ 40-50 Myr, \citealt{guirado_2011}), is  a member of the eponymous moving group AB Doradus Moving Group, associated with a group of 30 stars. It is an ultrafast rotator with a rotational speed v$\sin{i}$ = 91$\pm$1 km s$^{-1}$ \citep{cameron_2002} and a rotation period = 0.514 days \citep{kuerster_1994}. It has shown a high level of magnetic activity with an average $\mathrm{\frac{L_X}{L_{bol}}\sim10^{-3}}$. It is a quadruple system of a pair of binary stars AB~Dor~A, AB~Dor~B, 9.5$''$ away. AB~Dor~B is 60 times bolometrically fainter than AB Dor A. It is a binary system of AB~Dor Ba, AB~Dor Bb with 0.7$''$ separation first observed with Australian Telescope Compact Array at radio wavelengths \citep{Azulay_2015}. AB~Dor C is the close companion of AB~Dor A, with 0.16$''$ separation.

    \item BO Mic: An active K3V(e) star present in the constellation Microscopium, about 51.02 $\pm$ 1.5 parsecs away \citep{gaia_collab,gaia_EDR3}. It was first reported in the all-sky EUV ROSAT mission, with the largest stellar flare. It has a radial velocity of -6.5 $\pm$ 2.0 km s$^{-1}$ with an estimated v$\sin{i}$ of 134 $\pm$ 10 km s$^{-1}$ \citep{wolter_2005}; it is also known as Speedy Mic for its fast rotation \citep{bromage1992} with a rotation period of 0.380 $\pm$ 0.004 days \citep{cutispoto1997}. It is possibly a young pre-main-sequence star with high lithium abundance, strong photometric, and chromospheric activity \citep{Anders1993}. 
    It has been observed to have an average X-ray flux of 3.7 $\times$ 10$^{-12}$ ergs cm$^{-2}$ s$^{-1}$ \citep{Singh_1999}. Its differential rotation is weaker than the Sun, with its spot pattern changing nearly every 2.5 stellar rotations \citep{wolter_2005}.
    
    AB Dor and BO Mic are two prototypical active ultra-fast rotators which have been previously studied by us at X-ray wavelengths \citep{wolter_2008, lalitha_2013a, lalitha_2013b}. Both the stars show moderate flares in X-rays and/or UV in nearly every stellar rotation as well as occasional large flares. 

    \begin{table*}
	\centering
	\caption{Log of \emph{AstroSat} SXT and UVIT observations}
	\label{tab:tab2}
	\resizebox{2\columnwidth}{!}{
	\begin{tabular}{ccllllll} 
		\hline
	Star Name & Instrument & Observation ID & Start Time (UT) & Stop Time (UT) & Effective & Mean Count Rate & Energy band\\
        &   &    & Y:M:D:H:M:S & Y:M:D:H:M:S &  Exposure &  &\\
        &   &    &             &              & (s) &  &\\
		\hline	
	AB\,Dor & SXT  & 9000000274 & 2016:01:15:11:57:43 & 2016:01:16:04:46:40 & 17346 & 1.075$\pm$0.009& 0.3-6.0 keV\\
	AB\,Dor & SXT  & 9000000306 & 2016:01:31:11:13:32 & 2016:02:01:05:11:32 & 17272 & 1.132$\pm$0.009& 0.3-6.0 keV\\
    BO\,Mic & SXT  & 9000002104 & 2018:05:17:21:24:36 & 2018:05:20:12:34:13 & 41862 & 0.173$\pm$0.003& 0.3-6.0 keV\\
        & UVIT FUV & 9000002104 & 2018:05:17:22:42:12 & 2018:05:19:18:48:02 & 17138 & 0.872$\pm$0.134&1231-1731$\AA$\\
        &(F148W) & &&&&&\\
    DG\,CVn & SXT  & 9000001218 & 2017:05:11:07:47:57 & 2017:05:12:03:33:19 & 21687 & 0.065$\pm$0.002&0.3-6.0 keV\\
        &  UVIT NUV & 9000001218 & 2017:05:11:09:39:19 & 2017:05:12:03:10:48 & 81 & 9.391$\pm$3.121& 2026-2811$\AA$\\
        &  (N242W) &  &  &  &  & \\
        & UVIT FUV  & 9000001218 & 2017:05:11:09:39:14 & 2017:05:12:03:10:44 & 1102 & 0.396$\pm$0.164& 1231-1731$\AA$\\
        & (F148W) &  &  &&& &\\
    GJ\,3331 & SXT  & 9000000898 & 2016:12:18:13:56:41 & 2016:12:19:06:36:30 & 14748 & 0.200$\pm$0.004&0.3-6.0 keV\\
        &  UVIT NUV & 9000000898 & 2016:12:18:13:58:45 & 2016:12:19:09:59:29 & 15760 & 2.398$\pm$0.401&2747-2837$\AA$\\
        &  (N279N) &  & &&&&\\
        & UVIT FUV F148W & 9000000898 & 2016:12:18:13:58:49 & 2016:12:19:09:59:25 & 11749 & 0.316$\pm$0.124&1231-1731$\AA$\\
        & (F148W) & &&&&&\\
		\hline
	\end{tabular}}
	\newline
\end{table*}

    \item DG~CVn (GJ 3789): A binary system in which one of the components is a rapidly rotating M-type dwarf. DG~CVn is listed in the Washington Double Star Catalog as a binary of M dwarfs with separation $\sim$0.2$''$ located at a distance of 18.29 $\pm$ 0.11 parsecs \citep{gaia_collab, gaia_DR2}, with an estimated age of only 30 Myr. The system has fast rotation with $v\sin i \sim$ 55.5 km s$^{-1}$ and rotation period P$\mathrm{_{rot}}\sim$0.28d \citep{delfosse_1998, mohanty_basri_2003}. DG~CVn was detected in the ROSAT All-Sky Survey Bright Source Catalog with L$\mathrm{_X}$ $\sim 2.0\times 10^{28}$ ergs s$^{-1}$ and $\mathrm{\frac{L_X}{L_{bol}}} \sim -3.22$ \citealt{voges_1999}. It is also a radio-emitting source \citep{helfand_1999}. On 2014 April 23, \cite{drake_2014} reported a hard X-ray superflare on DG CVn, detected with the Burst Alert Telescope (BAT) onboard the {\it Swift} satellite, with a peak 15-50 keV flux of $\sim$300 mCrab \citep{osten_2016}. This flare was bright enough to trigger {\it Swift} and cause it to slew automatically to the source. This stellar super-flare is one of only a handful of stars that have been bright enough to trigger {\emph Swift} in such a fashion, and the previous ones \citep{osten_2007, osten_2010} have revealed the extreme end of magnetic reconnection in normal non-degenerate stars. 

    \item GJ 3331: A multiple stellar system, BD 211074, is reported in the Washington Catalog of Visual Double Stars as a stellar system consisting of three visual M-type stars \citep{mason_2001}. The primary component, GJ 3331, is an M1.5 star with V = 10.41 mag , the secondary component, GJ 3332 , an M2.5 star with V = 11.67 mag and the tertiary component, BD 211074C with a M5 or later spectral type \citep{jao_2003}. Our target of interest, GJ 3331, is a very rapidly rotating (P$\mathrm{_{rot}}$ $\sim$0.34d, \citealt{kiraga_2007}). However, \cite{messina_2014}
    found a P$\mathrm{_{rot}}$ $\sim$9.3d consistent with $v \sin i \sim 5.3$ km s$^{-1}$  \citep{reiners_2012, malo_2014}. It was detected in the extreme ultraviolet (EUV) by the \emph{ROSAT} Wide Field Camera (WFC) and listed in the WFC Bright Source Catalogue as RE 0506-213 \citep{pounds_1993}. GJ 3331 is an active stars with a strong X-ray emission detected by \emph{ROSAT} ( L$\mathrm{_X} \sim 1.4 \times 10^{29}$ ergs s$^{-1}$ and $\mathrm{\frac{L_X}{L_{bol}}} \sim -3.01$, \citealt{voges_1999}).  Both components, A and the unresolved BC, are listed in the catalog of UV~Ceti-type flare stars compiled by \cite{gershberg_1999}.

\end{enumerate}

This paper is organised as follow: in \S 2  we describe the \emph{AstroSat} observation in both X-rays and UV wavelengths and the data reduction. In \S 3, we present our data analysis and results from the temporal and spectral analysis. A detailed discussion for each of the target stars and our conclusion is presented in \S 4.

\begin{figure*}
\begin{center}
\includegraphics[width=0.3\textwidth,clip]{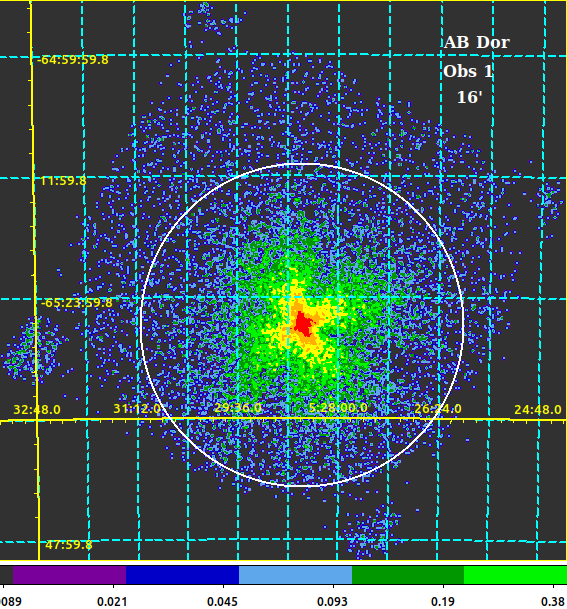}
\hspace{0.2em}
\includegraphics[width=0.3\textwidth,clip]{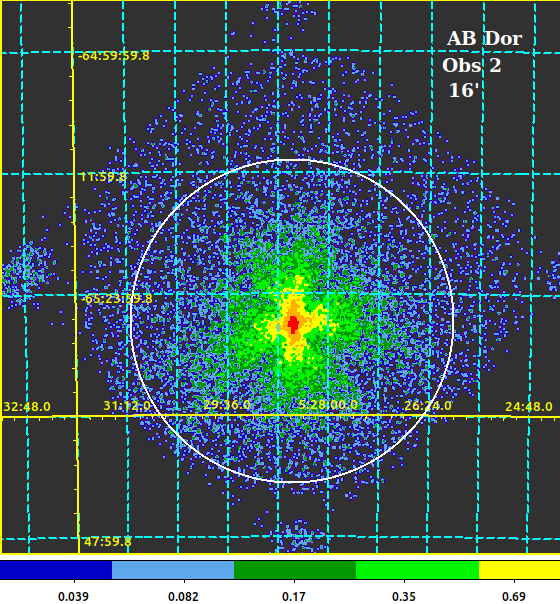}
\hspace{0.2em}
\includegraphics[width=0.315\textwidth,clip]{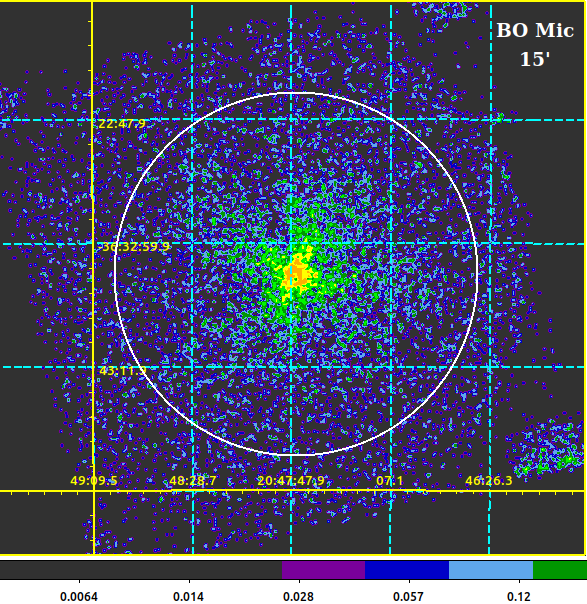}
\hspace{0.2em}
\vspace{1em}
\includegraphics[width=0.3\textwidth,clip]{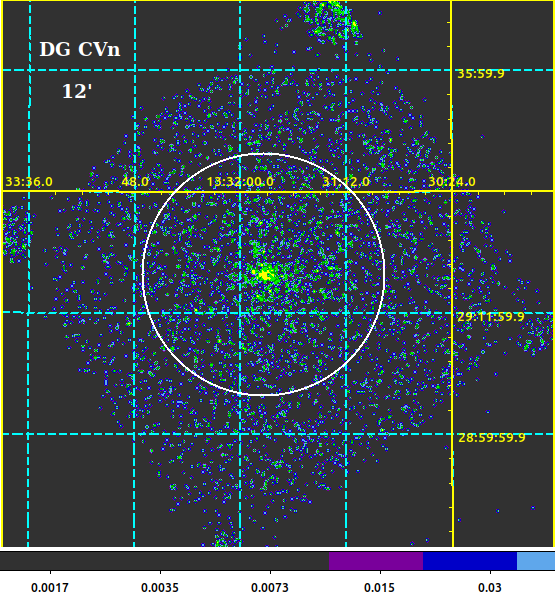}
\hspace{0.2em}
\includegraphics[width=0.3\textwidth,clip]{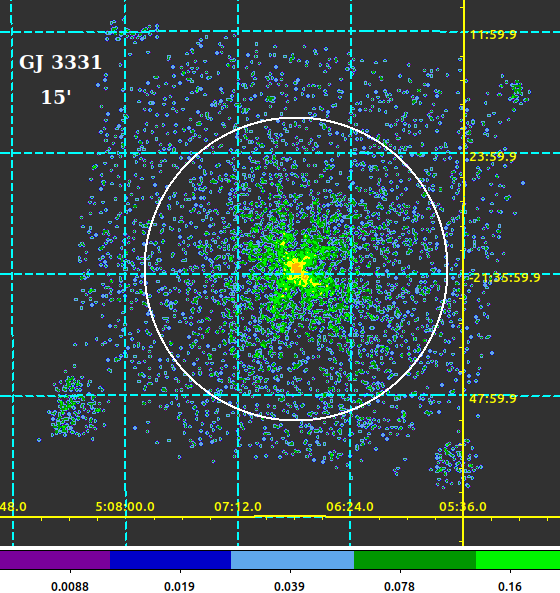}
\caption{\label{img} X-ray images of the stars in 0.3-3.0 keV energy band observed with the {\it AstroSat} SXT (Clockwise from top left: AB~Dor (1st epoch), AB~Dor (2nd epoch), BO~Mic, DG~CVn and GJ~3331). The white circle shows the extraction region for the light curves and spectra and the image has been smoothed with a Gaussian function of 3-pixel radius.}
\end{center}
\end{figure*}

\section{ Observation and data reduction}

The targets were observed with \emph{AstroSat} \citep{singh_2014}. 
\emph{AstroSat} is a multi-wavelength astronomy mission carrying five multi-band payloads. The stars listed here were observed with 
the SXT covering the 0.3-6.0 keV energy band \citep{singh_2016,singh_2017}, and the Ultra-Violet Imaging Telescope \citep{Tandon_2017, Tandon_2020}.  The UVIT consists of two 35cm Ritchey-Chrètien telescopes, one covering the FUV (1250 – 1830 Å) band and the other telescope uses a dichroic beam splitter that splits the beam into the  NUV (1900 – 3040 Å) and the visible (VIS: 3040 – 5500 Å) channels.  Each waveband has a choice of filters as given in \citep{Tandon_2017}, and have $\sim1.8''$ resolution. The CMOS imagers used for all the bands have 512 x 512 pixels, with each pixel mapped to 8 $\times$ 8 sub-pixels, giving a pixel scale of 0.416 arc-seconds/sub-pixel \citep{Tandon_2020}. A detailed description of the observations is given in Table~\ref{tab:tab2}.

\begin{figure*}
\begin{center}
\includegraphics[width=0.3\textwidth,clip,trim={0 0 0 0.85cm}]{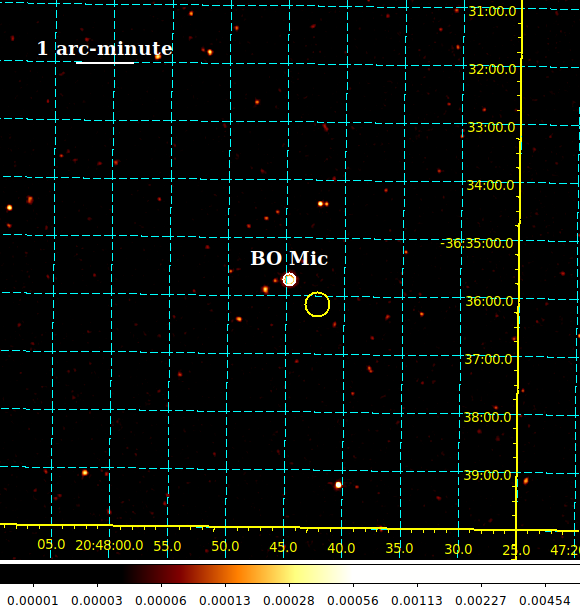}
\hspace{0.3em}
\includegraphics[width=0.3\textwidth,clip]{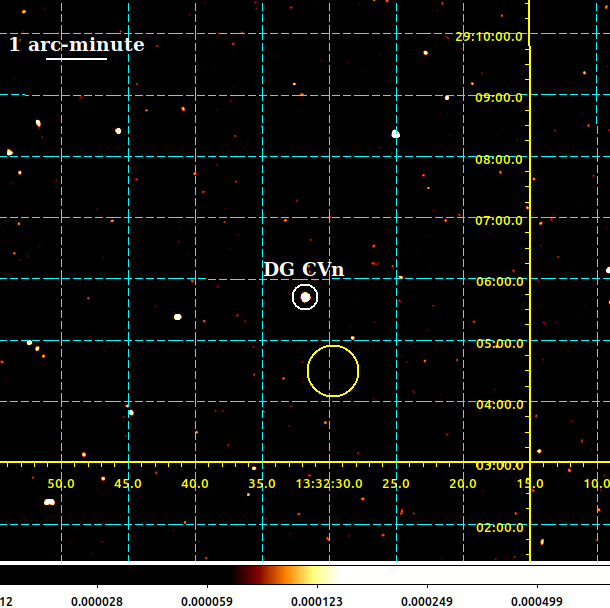}
\hspace{0.3em}
\includegraphics[width=0.28\textwidth,clip,trim={0 0 0 0.47cm}]{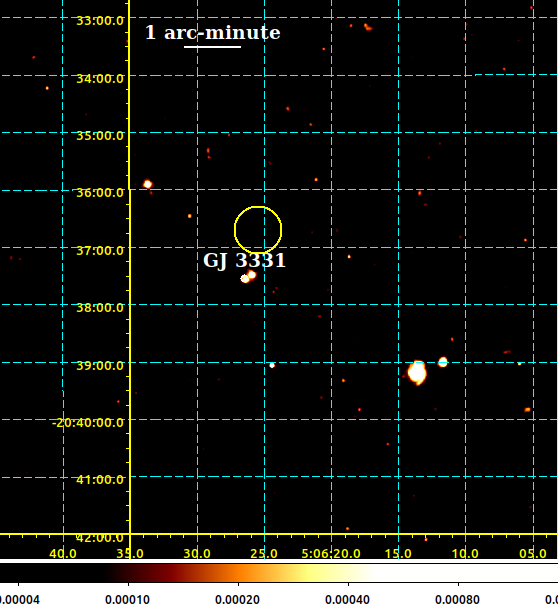}
\caption{\label{fig:fuv} FUV images for each observation with the \emph{AstroSat UVIT} (BO Mic F148W, DG CVn F148W, GJ 3331 F148W, respectively). The white circle shows the extraction region for the light curves and the yellow circle shows the background region. The images have been smoothed with a Gaussian function of 5-pixel radius, while the DG CVn F148W image from orbit 3 has been smoothed with a Gaussian function of 11-pixel radius due to the absence of combined orbits image.}
\end{center}
\end{figure*}

\subsection{Soft X-ray Telescope (SXT)}
X-ray data from individual orbits (level 1 data) were reduced with the standard pipeline for the SXT using SXTPIPELINE software at the Payload Operation Centre (POC) where events are selected with event grading similar to \emph{Swift}-XRT, see \citet{romano_2005} and events with grades $>$ 12 are removed.  All events above a preset threshold above the electronic noise level were time-tagged, applied with the coordinate transformation from the detector to sky coordinates, and bias subtracted. The bad pixels tagging, search and removal of hot and flickering pixels, and conversion from the event pulse height to the X-ray energy of the event were carried out in the pipeline.  The selected events are screened for bright Earth avoidance angle of $\ge$ 110 degrees and those detected during the passage through the South Atlantic Anomaly (SAA) were removed using the information obtained from the Charged Particle Monitor (CPM) ensuring that the CPM rate is below 12 counts s$^{-1}$.   Good Time Intervals (GTI) were thus generated and applied, producing level 2 data events files.  A merged events file of all such cleaned events from GTIs was generated using a Julia script provided by the SXT POC team. Any residual contamination from non-X-ray events was further  removed by examining the light curves for flares or dips in the energy range of 7 - 10 keV or by looking at simultaneously observed  background regions if available. The useful exposure times from merged cleaned events files are listed in Table 2 for each observation.

X-ray images are made for the cleaned events in the energy range of 0.3 - 3.0 keV for AB~Dor Obs 1 (id:274), AB~Dor Obs 2 (id:306), BO~Mic, DG~CVn and GJ~3331. The images produced have been smoothed by a Gaussian with a kernel of 3 pixels (1 pixel = 4$''$) radius. Source counts for light curves and spectra were extracted from a circular region shown by white circles in Figure \ref{img}. The extraction radius for AB~Dor Obs 1 \& 2 was 16$'$, whereas for BO Mic \& GJ 3331 it was 15$'$, and 12$'$ for DG~CVn. The extraction radius was decided after examining the radial profile of surface brightness using 25 annuli from 0.5$'$ to 19$'$ and making sure that $\geq$ 95\% of all the source counts were selected considering the large point spread function (PSF) of the SXT \citep{singh_2017}.  The background was obtained from an annular region of 13-18$'$ for DG~CVn, while the background for other sources was taken from other observations of a deep sky region provided by the SXT POC.

The X-ray counts in the spectra of sources extracted were grouped using the {\it grppha} tool to ensure a minimum of 30 counts per bin. The background spectra were extracted using annuli from 13$'$ to 18$'$ for weaker sources like DG~CVn and a deep-sky background file has been used for the rest of the observations. 
The response matrix (sxt\_pc\_mat\_g0to12.rmf) and standard ARF files (sxt\_arf\_excl00\_v04\_20190608.arf) used are available at the SXT POC website
\footnote{\href{http://www.tifr.res.in/~astrosat_sxt/index.html}{http://www.tifr.res.in/$\sim$astrosat$\_s$xt/index.html}}.

\subsection{Ultra Violet Imaging Telescope (UVIT)}
The UVIT data were analysed using the Level 2 photon lists and images obtained from the ISSDC. The images were compared with Galaxy Evolution Explorer (GALEX) all-sky imaging survey (AIS) images for FUV and NUV respectively. The target stars were identified by matching stellar patterns in SAOImage DS9,
  and the astrometric corrections were done by translating and rotating the coordinates through \emph{astropy} version 4.0.1. The astrometric calibrations were also verified with the SIMBAD catalog.
The astrometric error achieved for BO~Mic F148W (1231 - 1731 $\AA$) field is 3$''$ (Figure~\ref{fig:fuv},\ref{fig:nuv}), DG~CVn F148W is 4$''$, DG~CVn N242W  (2026 - 2811 $\AA$) is 5$''$ (Figure~\ref{fig:fuv},\ref{fig:nuv}), GJ~3331 F148W is 2$''$, and GJ~3331 N279N (2747 - 2837 $\AA$) is 2$''$ (Figure~\ref{fig:nuv}) from the SIMBAD catalog.

The light curves extraction was done using the photon list file for each orbit through the \emph{curvit} python package \citep{curvit}. The target star for each orbit is found by matching stellar patterns and by transforming astronomical coordinates to instrumental coordinates. The FUV PSF pedestal sets the minimum source radius for extraction, i.e., 3.33$''$ \citep{Tandon_2017}, whereas the maximum source radius is 39.52$''$ \citep{Tandon_2020}. The source radius was chosen based on count rate values variation on changing source radii while ensuring no other event contributes to the count rate values. The background radius was twice the source radius in most cases. The background region was kept within one arc-minute of the source with no event inside it. The time bin width was chosen based on the exposure time for that observation. Different time bin widths were compared to achieve a sufficient number of data points with small errors for each data point. The details for each extraction  are as follows:

\begin{figure*}
\begin{center}
\includegraphics[width=0.38\textwidth,clip]{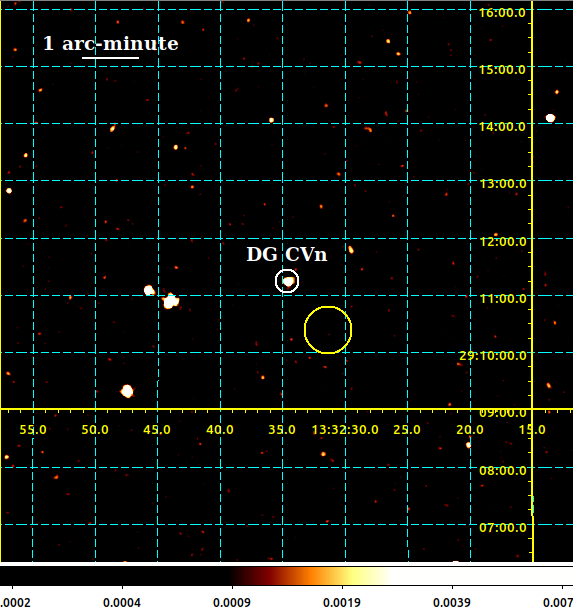}
\hspace{1em}
\includegraphics[width=0.395\textwidth,clip]{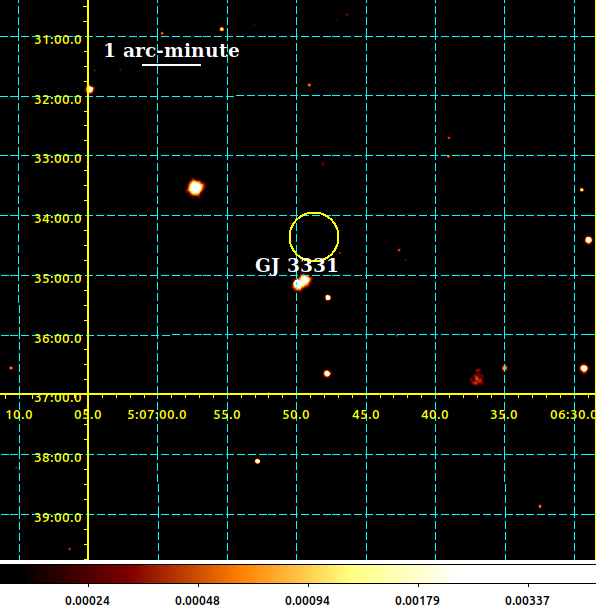}
\caption{\label{fig:nuv} NUV images for each observation with the \emph{AstroSat} UVIT (GJ 3331 N279N, and DG CVn N242W). The white circle shows the extraction region for the light curves and the yellow circle shows the background region. The images has been smoothed with a Gaussian function of 5 pixel radius, while DG CVn F148W image from orbit 3 has been smoothed with a Gaussian function of 11 pixel radius due to absence of combined orbits image.}
\end{center}
\end{figure*}

\begin{figure*}
\begin{center}
\includegraphics[width=0.49\textwidth,clip]{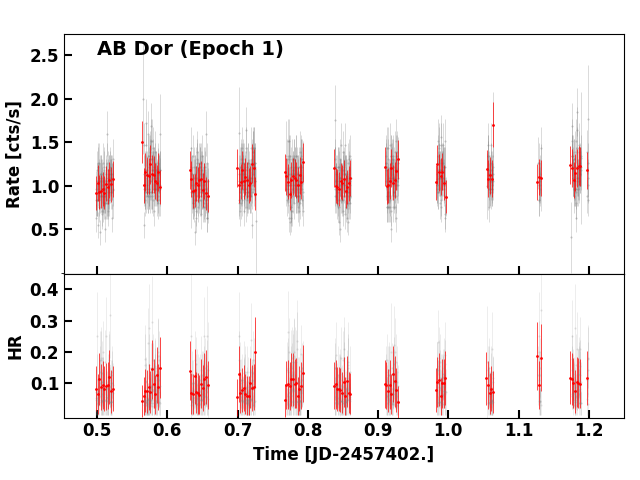}
\includegraphics[width=0.49\textwidth,clip]{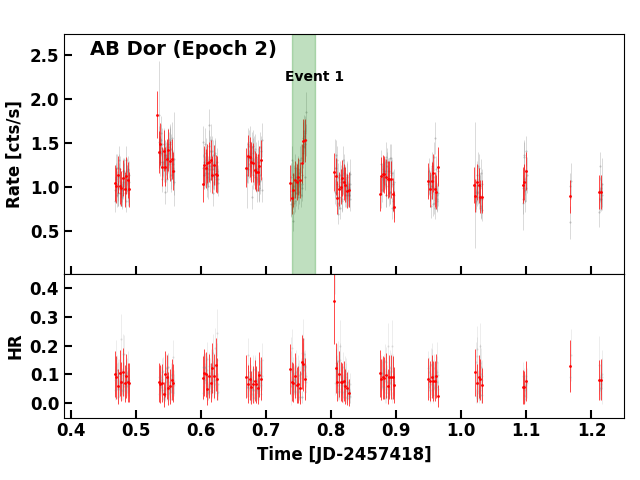}
\includegraphics[width=0.49\textwidth,clip]{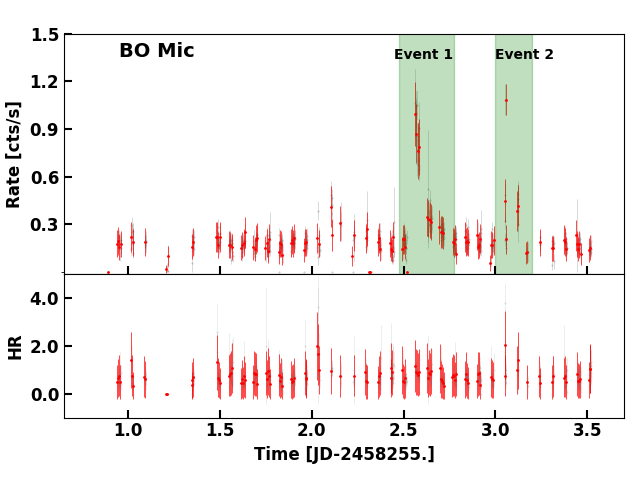}
\includegraphics[width=0.49\textwidth,clip]{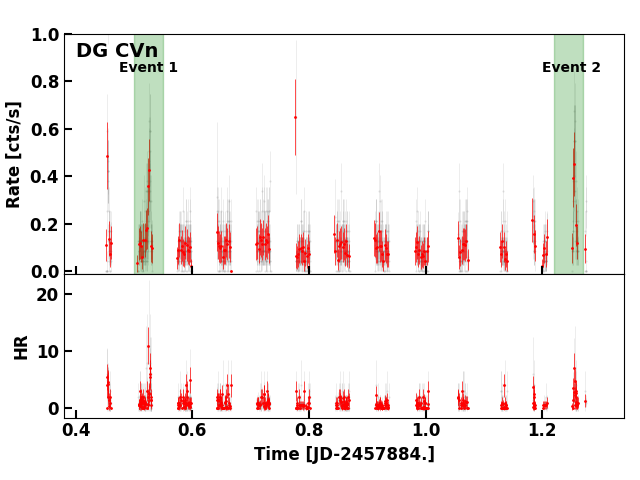}
\includegraphics[width=0.49\textwidth,clip]{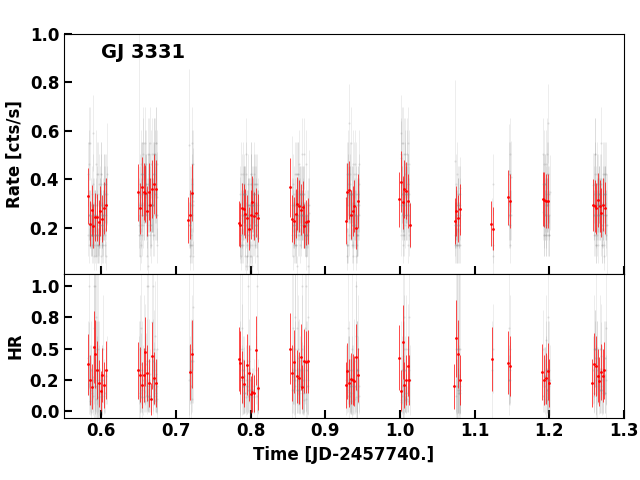}

\caption{\label{fig:lc1} X-ray light curves and their corresponding hardness ratios are plotted for each of the targets. Over-plotted in red is the moving averages for better visualisation of the associated variability. The time segments 
corresponding to the flare-like events are represented by the vertical green blocks.}
\end{center}
\end{figure*}

\begin{itemize}

  \item BO~Mic has a 12.5$''$ source radius, 25.0$''$ background radius, and a 50s wide time bin. A flare is observed near the end of the observation in the UVIT-F148W light curve. 
  \item DG~CVn has a 12.5$''$ source radius, a 25.0$''$ background radius, and a 15 seconds wide time bin for both F148W \& N242W filters. 
  \item GJ~3331 has a smaller radius 3.33$''$ for extracting the source events due to the proximity of a secondary member GJ~3332, a 25.0$''$ background radius, and a 20s wide time bin for F148W data, and 15s wide time bin for N279N data.
\end{itemize}

\section{Data analysis and results}

\subsection{X-ray Temporal analysis}
The X-ray light curves for our four targets were made in three energy bands - soft, hard and total for all the sources. The total energy band used is 0.3-6.0 keV, whereas the soft and hard energy bands were guided by the spectral analysis that follows to have some physical meaning relative to the spectral components identified therein and due to the presence of significant flaring episodes in some of them.  

The X-ray light curves for our four targets were made in three energy bands - soft, hard and total for all the sources. The total energy band used is 0.3-6.0 keV, whereas the soft and hard energy bands were influenced by the subsequent spectral analysis. These choice were made to ensure that these energy bands have a meaningful connection to the spectral components identified in the analysis. Furthermore, the decision was influenced by the presence of significant flaring episodes observed in some of the sources. By focusing on these specific energy bands, we aim to gain insights into the physical mechanisms behind the X-ray emissions, enabling us to better understand the behaviours and characteristics of the sources.  These bands and the corresponding figures shown here are: 
\begin{itemize}
  \item AB~Dor: 0.3-2.0 keV for soft X-ray band and 2.0-6.0 keV for hard X-ray band. 
  \item BO~Mic: 0.3-1.2 keV for soft X-ray band and 1.2-6.0 keV for hard X-ray band.
  \item DG~CVn: 0.3-1.0 keV for soft X-ray band and 1.0-6.0 keV for hard X-ray band.
  \item GJ~3331:0.3-2.0 keV for soft X-ray band and 2.0-6.0 keV for hard X-ray band. 
  \end{itemize}

 In Figure~\ref{fig:lc1} (top panel), we plot the background-subtracted SXT light curves for AB~Dor, BO~Mic, DG~CVn and GJ~3331. In Figure~\ref{fig:lc1} (bottom panels), the hardness ratios (HR) derived from the ratio of counts in the hard band divided by the counts in the soft are plotted for each of our target. Three of our targets AB Dor, BO Mic and DG CVn show several flaring events represented by green vertical blocks in Fig.\ref{fig:lc1}. 
  
 During the observations several flare-like events were detected in our targets. AB~Dor exhibited a flare-like event approximately 25~ks into the observations during Epoch 2. The count rate in the SXT light curves in the 0.3-6.0 keV range increased from a quiescent value of approximately 0.9 cts~s$^{-1}$ to 1.6 cts~s$^{-1}$.

BO~Mic showed two flare-like events. The first event occurred at approximately 145ks ($\sim$JD-2458257.55) into the observations, with the quiescent flux increasing from 0.25 cts~s$^{-1}$ to 1.05 cts~s$^{-1}$. The second event occurred at approximately 180~ks ($\sim$JD-2458258.1), with the quiescent flux increasing to 1.2 cts~s$^{-1}$.

Similarly, DG~CVn displayed multiple flare-like events. The first event occurred at approximately 7ks (JD-2457884.55) into the observations, with the count rate increasing from 0.12 cts~s$^{-1}$ to 0.55 cts~s$^{-1}$. Additionally, event 2 occurred at JD-2457885.22 showing similar changes in count rate as event 1. In Figure~\ref{fig:lc1} for DG~CVn, we note heating related trends in the HR. However, due to limitations in the SNR, conducting a detailed spectral analysis for these heating events observed in DG~CVn may not be feasible at this time. 

On the other hand, no significant flaring episodes were detected in GJ~3331 and during the first epoch observations of AB~Dor. Furthermore, although slow variations on timescales of several hours could be observed in the light curves, we note a lack of any significant signal attributed to rotation modulation.

\begin{figure}
\begin{center}
\includegraphics[width=0.47\textwidth,clip]{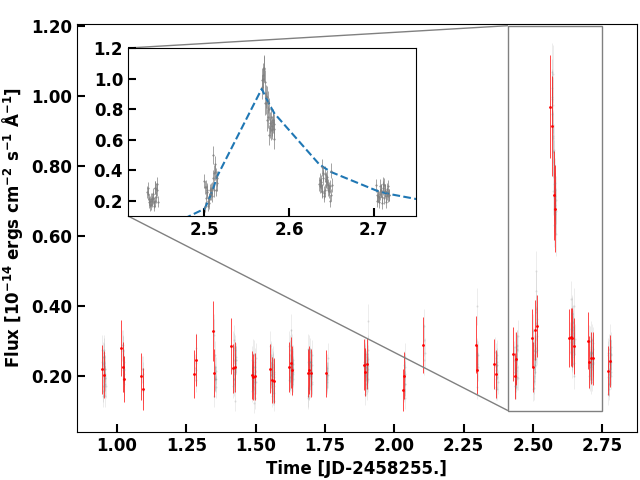}

    \caption{\label{fig:uvit_bomic}  Combined orbits FUV light curve for BO Mic binned to 50s. }

\end{center}
\end{figure}

\subsection{UV Temporal analysis}

In the following, we examine the UV emission for three stars with FUV observations and two stars with NUV observations. UV fluxes were calculated by multiplying the unit conversion factor to the measured counts per second for each filter \citep{Tandon_2020}. The measured FUV and NUV fluxes for BO Mic and GJ 3331 are listed in Table~\ref{tab:uvflux}.

\begin{figure*}
\begin{center}
\includegraphics[width=0.47\textwidth,clip]{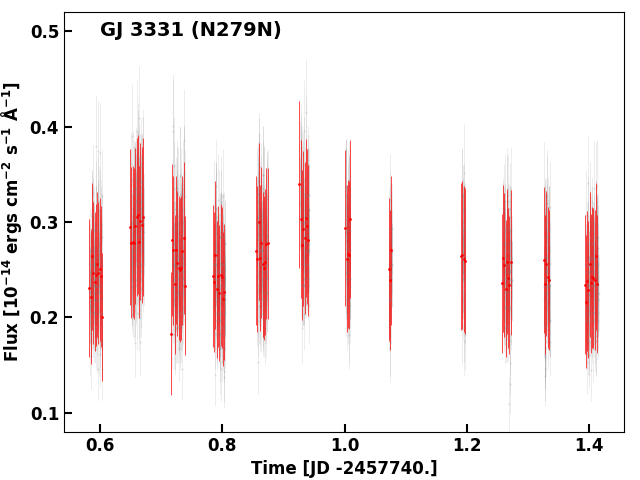}
\includegraphics[width=0.47\textwidth,clip]{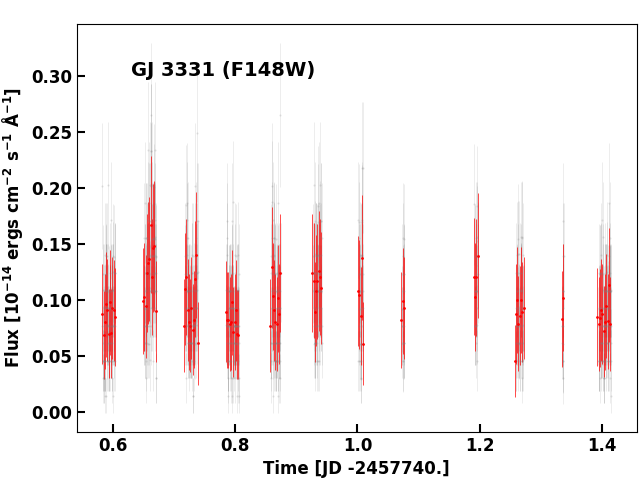}
    \caption{\label{fig:gj3331_uvit}  Combined orbits light curves in NUV (left panel) and FUV (right panel) for GJ 3331. The bin widths used are 15s and 20s for NUV and FUV, respectively. For better visualisation of the variability we over plot a  moving average light curve in red. }
\end{center}
\end{figure*}

\begin{figure}
\begin{center}
\includegraphics[width=0.47\textwidth,clip]{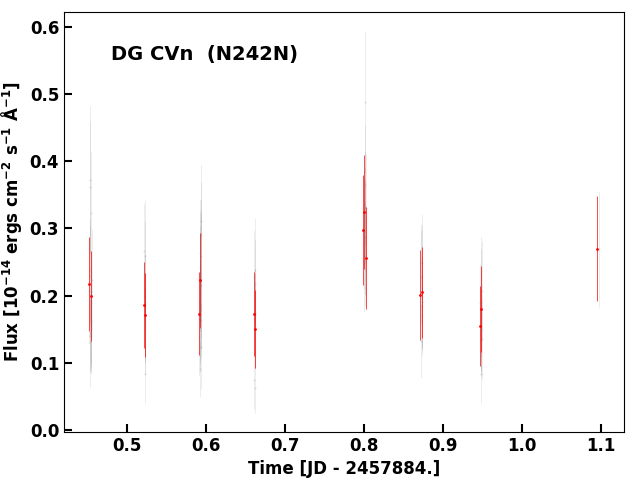}
\includegraphics[width=0.47\textwidth,clip]{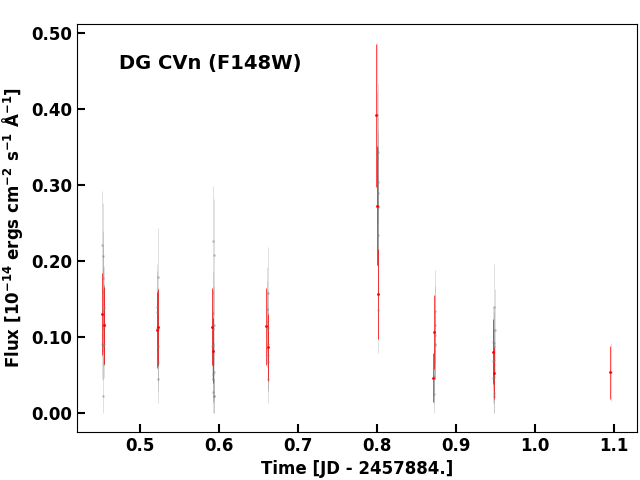}

    \caption{\label{fig:dgcvn_uvit}  Combined orbits light curve in NUV (top panel) and FUV (bottom panel) for DG CVn binned to 15s. A flare-like event at JD$\sim$2457884.079 (see Fig~\ref{fig:lc1}).}

\end{center}
\end{figure}

\begin{itemize}
 \item \noindent BO Mic: 
    In Figure \ref{fig:uvit_bomic}, we plot the combined orbit FUV light curve for BO Mic observed with F148W filter and binned to 50s. We observe the large flare-like event (event 1) in both X-rays and FUV. The flux increases from 0.2$\times 10^{-14}$ to 1.02$\times 10^{-14}$ ergs cm$^{-2}$ s$^{-1}$ $\AA^{-1}$ in FUV.

    \item \noindent GJ 3331: In Figure \ref{fig:gj3331_uvit}, we plot the combined orbit light curves for NUV band observed with N279N filter (left panel) and FUV with F148W filter (right panels). The average NUV and FUV fluxes is 2.5$\times 10^{-15}$ and 1.1$\times 10^{-14}$  ergs cm$^{-2}$ s$^{-1}$ $\AA^{-1}$, respectively. Although the light curve shows small scale modulation we were unable to obtain any significant periodicities associated with the modulation.

  \item \noindent DG CVn: In Figure \ref{fig:dgcvn_uvit}, we plot the combined orbit NUV light curve observed with N242W filter (top panel) and FUV light curve (bottom panel) observed F148W filter. Both the light curves are binned to 15s. A flare-like event observed in X-ray (Event 2 in Figure~\ref{fig:lc1}) which occurred around MJD$\sim$2457884.79 was also covered by both the UV filters. During this event the flux increased from the quiescent value of  $\sim 0.13 \times 10^{-14}$  to $\sim 0.41 \times 10^{-14}$  ergs cm$^{-2}$ s$^{-1}$ $\AA^{-1}$ in NUV and FUV bands.
    The flux changes by a factor of  3.2, however, we do not resolve the rise and decay phase of the event. Although we observe X-ray excess in the SXT light curve (DG CVn Event~3 Fig~\ref{fig:lc1}), we believe the flare rise and decay information is lost due to Earth block in UV and X-rays, respectively. 
    
\end{itemize}

\subsection{X-ray Spectral Analysis}

\begin{figure*}
\begin{center}
\includegraphics[width=0.49\textwidth,clip]{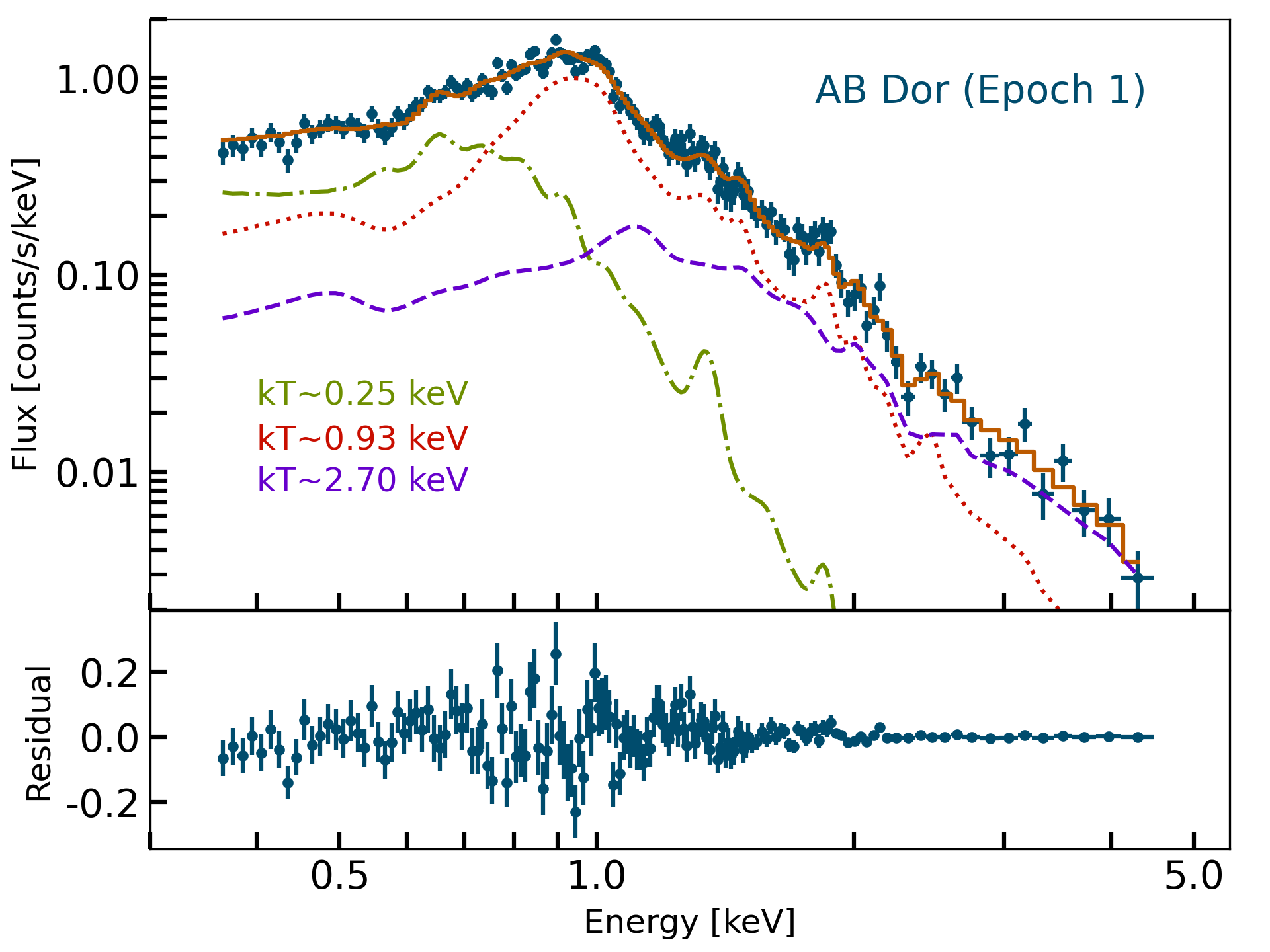}
\includegraphics[width=0.49\textwidth,clip]{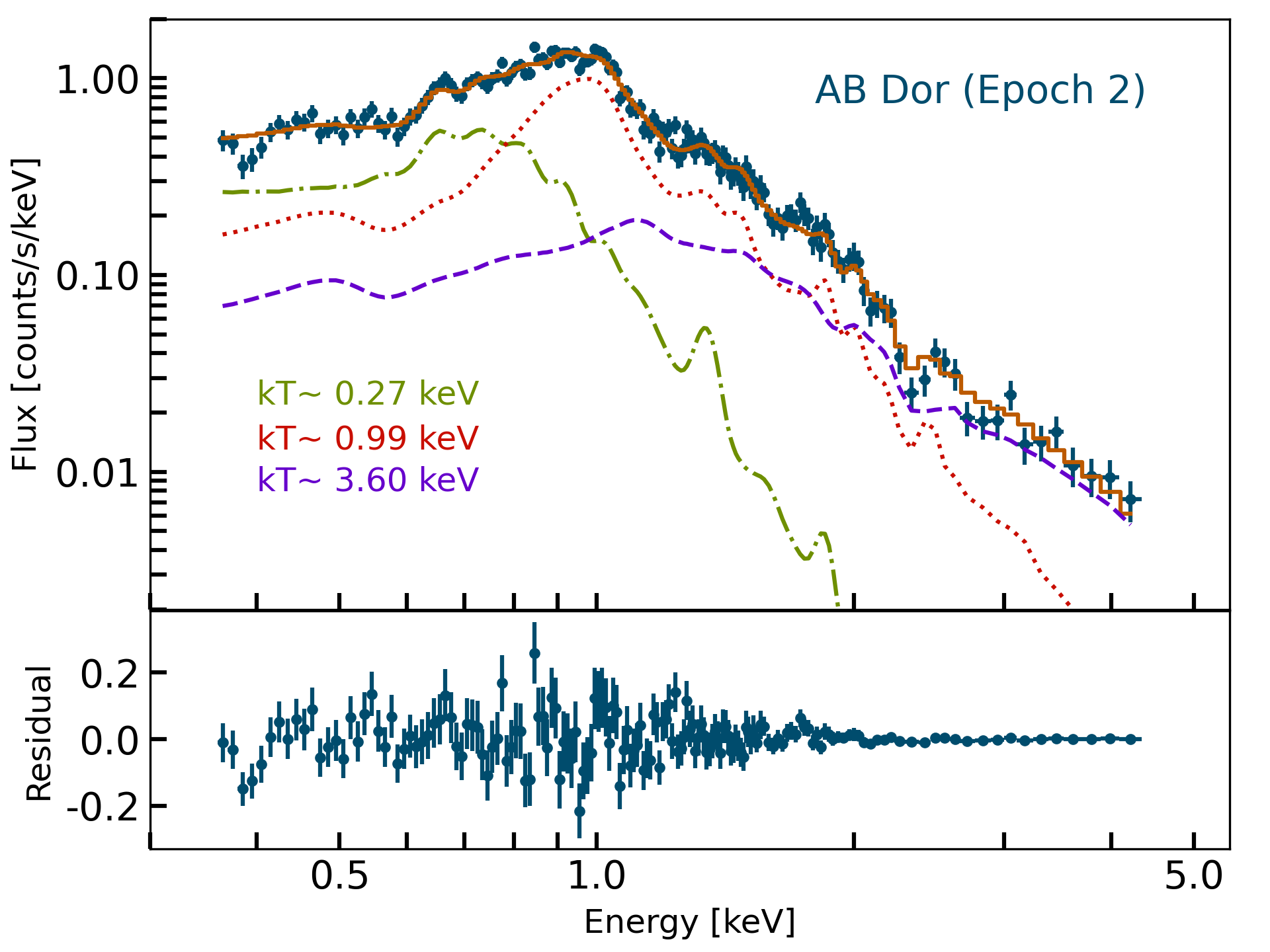}
\includegraphics[width=0.49\textwidth,clip]{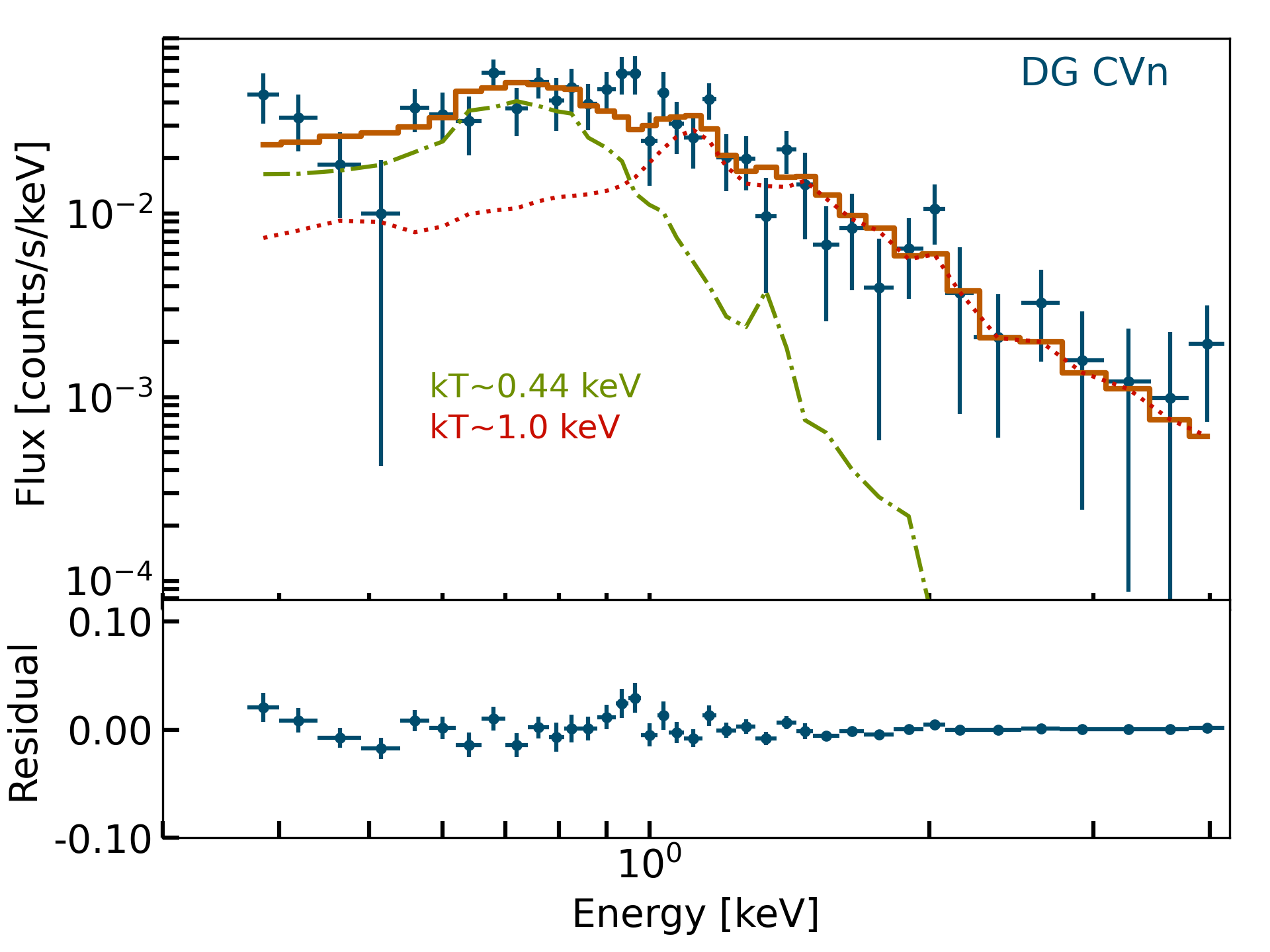}
\includegraphics[width=0.49\textwidth,clip]{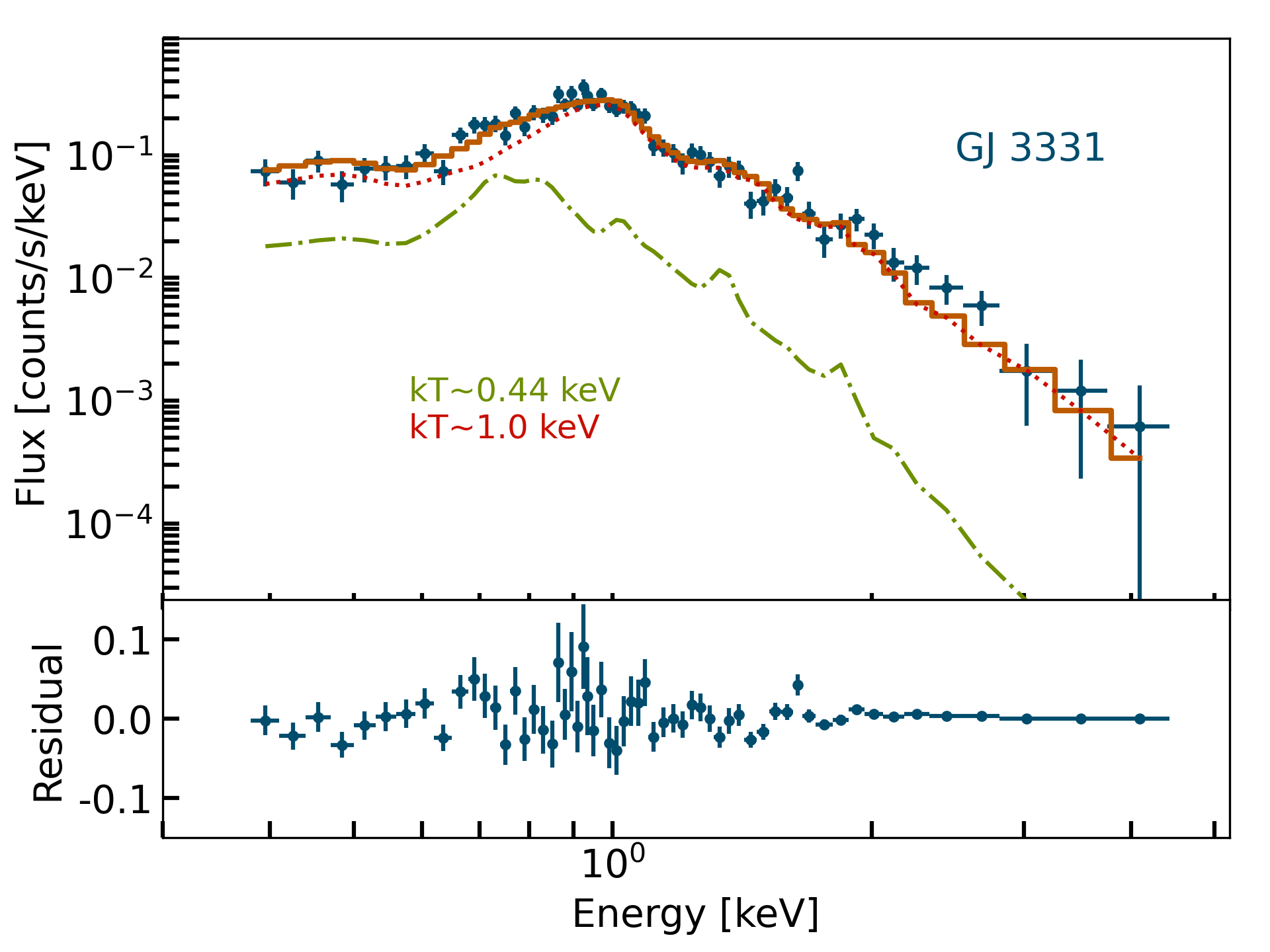}
\caption{\label{fig:specall} X-ray spectra of AB Dor (two epochs), DG~CVn and GJ~3331 along with the best-fit models as observed with  \emph{AstroSat} SXT.}
\end{center}
\end{figure*}

\begin{figure*}
\begin{center}
\includegraphics[width=0.49\textwidth,clip]{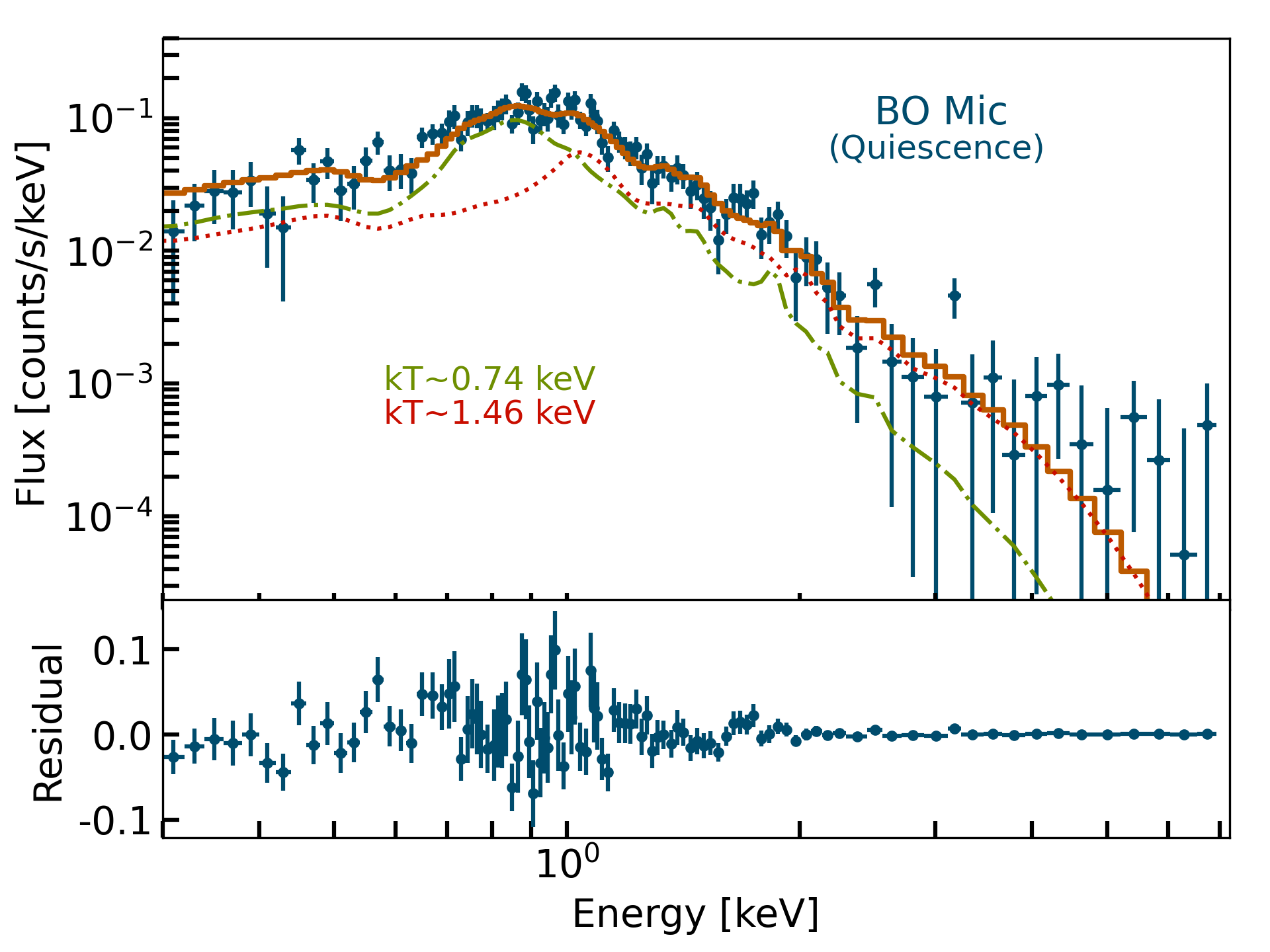}
\includegraphics[width=0.49\textwidth,clip]{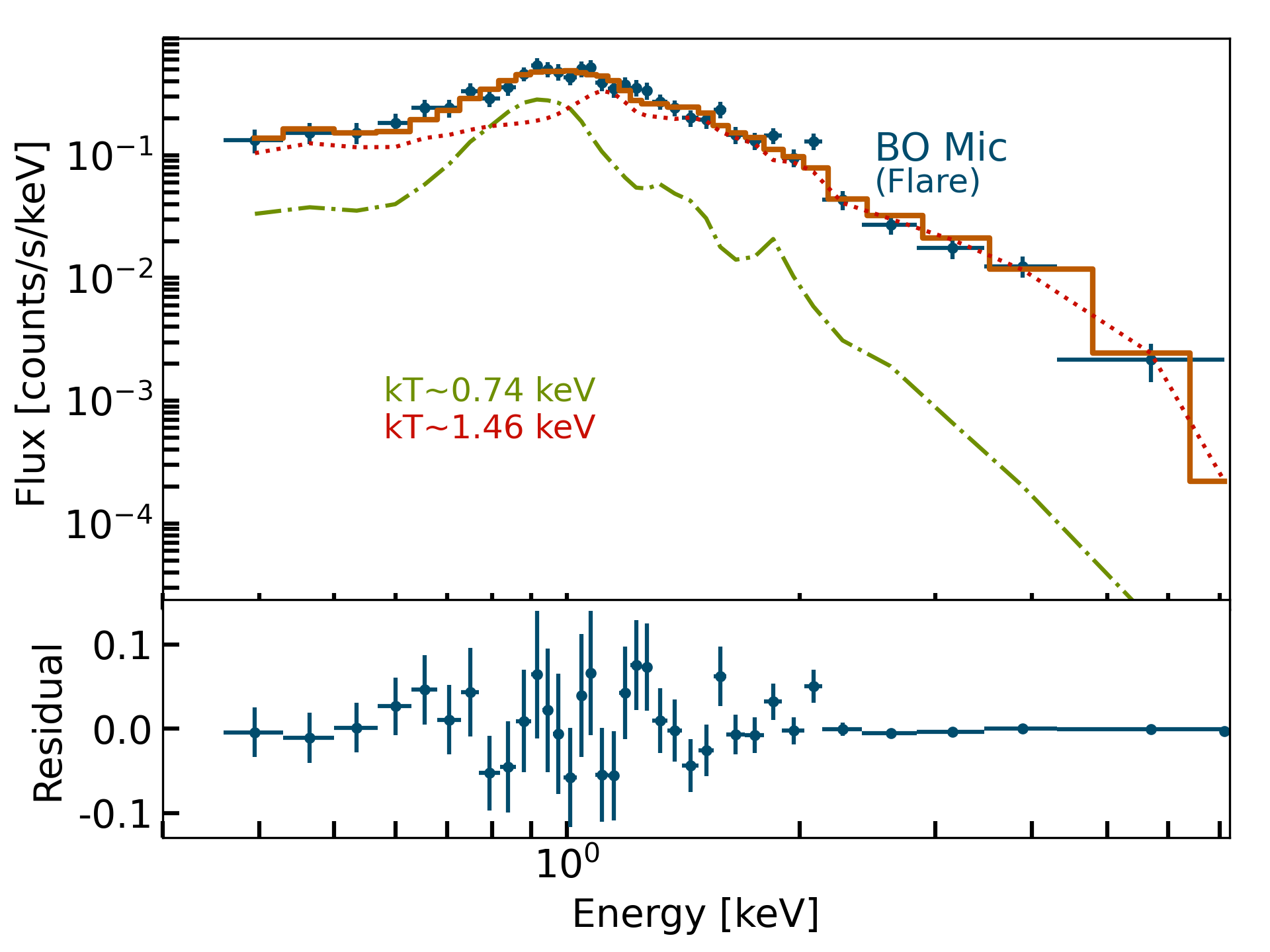}
\caption{\label{fig:specbomic} X-ray Spectra of BO~Mic and best-fit models as observed with  \emph{AstroSat} SXT during the quiescent state (left) and flaring state (right).}
\end{center}
\end{figure*}

Spectral analysis was carried out with XSPEC version 12.9.1. \citep{xspec} distributed with the heasoft package (version 6.20). The spectra were fitted with optically-thin plasma emission models known as Astrophysical Plasma Emission Code ({\it apec}) 
as described by \cite{smith_2001} using the atomic database AtomDB version 3.0.7 
\footnote{\href{http://www.atomdb.org}{http://www.atomdb.org}}. 
The spectral data used for fitting were restricted to 0.35-4.5 keV due to uncertainty in the background subtraction above 4.5 keV. The SXT spectra for all our targets are shown plotted in Figure~\ref{fig:specall}  and Figure~\ref{fig:specbomic}.

\begin{table}
\centering
\caption{Measured FUV and NUV fluxes}
\label{tab:uvflux}
\begin{tabular}{lcc} 
\hline\hline
Star name  &  FUV   & NUV \\ [0.5ex] 
\hline
BO Mic (quiet) & 2.1$\pm$0.4& --\\
BO Mic (flare) & 7.8$\pm$0.5&\\
DG CVn&1.2$\pm$0.5& 2.0$\pm$0.7\\
GJ 3331&1.1$\pm$0.3&2.5$\pm$1.1\\

\hline
\end{tabular}
\begin{tablenotes}

Note: All errors quoted are with 90\% confidence.\\ 
Fluxes in 10$^{-15}$ ergs cm$^2$ s$^{-1}$ were calculated using the unit conversion (UC) factor x 10$^{-15}$, where UC is: 3.09 $\pm$ 0.03 for F148W, 0.22 $\pm$ 0.001 for N242W, 3.50 $\pm$ 0.04 for N279N \citep{Tandon_2020}.
\end{tablenotes}

\end{table}

To account for the absorption of X-ray photons by interstellar material, a multiplicative absorber model called {\it Tbabs} was employed.  
The equivalent Galactic neutral hydrogen column density (N$\mathrm{_H}$) was fixed at a low value of 10$^{20}$ cm$^{-2}$. 
The elemental abundance table {\it aspl}   \citep{As2009} was used in our analysis.

We used the $\chi^2$ minimisation technique to find the best-fit 
parameters of the plasma emission models. 
We present the results of our final best fitting model, including the values of the parameters derived, along with their 90\% confidence error range in Table~\ref{tab:Tab3}. Various model combinations were examined to arrive at the best-fitting models, and the corresponding results are provided in Table~\ref{tab:Tab-appendix}. We started with assuming a single temperature model {\it apec} with solar abundances, and then progressed to non-solar abundances before trying additional temperature components and thus converging to the final best-fit models shown in Table~\ref{tab:Tab3}. The elemental abundances for all the temperature components were assumed to be the same in each case.
We find that two temperature component models with sub-solar abundances lead to acceptable best fits and provide an adequate description of the data for three stars: BO~Mic, DG~CVn and GJ331. AB~Dor, however, required a minimum of three temperature components with sub-solar abundances.

During the quiescence and the flare observed from BO Mic, we carried out a quantitative study of plasma temperatures and global abundance values (Figure~\ref{fig:specbomic}). The plasma emission measure increases with an increase in the temperature. We also notice that the spectra harden as the temperature increases. Furthermore during the flare, the global elemental abundance increases indicating that fresh chromospheric material is transported into the corona changing the elemental abundance temporarily.

\begin{figure*}
\begin{center}
\includegraphics[width=0.49\textwidth,clip]{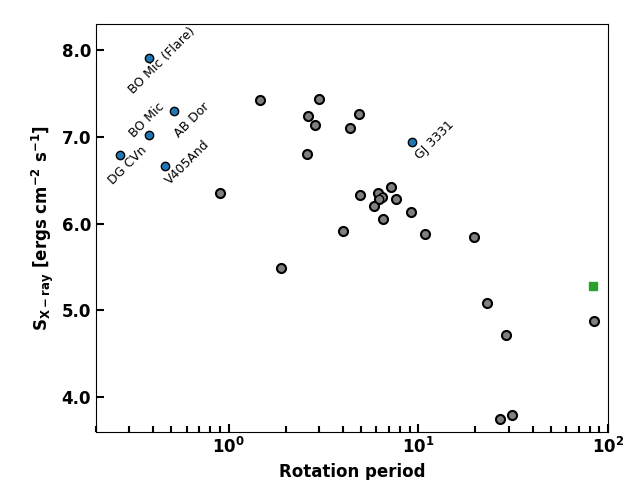}
\includegraphics[width=0.49\textwidth,clip]{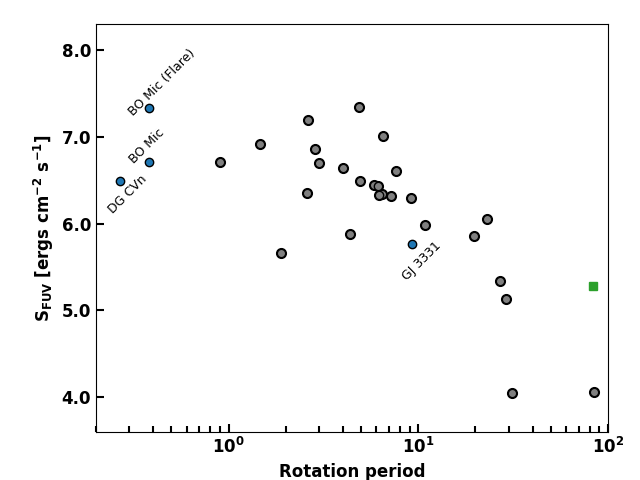}
\caption{\label{fig:lum} The soft X-ray (left panel) and FUV (right panel) surface fluxes are plotted as a function of the rotation period.  The stars studied in this work are depicted as filled circles and open circles depict the stars from \citep{france_2018}. Represented as green square is Proxima Centauri observed previously with \emph{AstroSat} \citep{lalitha_2020} and HST \citep{youngblood_2017}. }
\end{center}
\end{figure*}

\section{Discussion and Conclusion}

In this work,  we have analysed X-ray and far-UV data from \emph{AstroSat} observations of four ultra-fast rotating active stars AB~Dor, BO~Mic, DG~CVn and GJ~3331. These observations provide a valuable insight into X-ray and UV emission of these stars, allowing us to investigate their outer atmospheric properties and activity levels. 

\begin{enumerate}
    \item  AB Dor: We found the coronal parameters remained consistent between the two epochs. The spectral analysis revealed a three-temperature model with temperatures between 0.2-1.0 keV (3-12 MK). In \citep{lalitha_2013a}, a fixed four-temperature VAPEC model with KT values of 0.3, 0.6, 1.2, and 2.4 was used, resulting in emission measures ranging from 1.43 to 4.59E52 cm$^{-3}$. However, due to lower SNR in our data, we employed a three-temperature APEC model instead, allowing the models to vary freely while keeping the abundance fixed across components. The derived temperatures were in the range of 0.25-3.6 keV, with emission measures spanning from 2.2 to 6.4$\times$52 cm$^{-3}$. Importantly, the measured emission measures from our new work are consistent with those reported in the previous study by \citep{lalitha_2013a}.  The soft X-ray flux obtained for AB~Dor during the two epochs are consistent with the X-ray activity cycle predicted for AB~Dor (\citealt{lalitha_2013b}), indicating that the star is heading towards activity maxima. The observed X-ray and FUV fluxes of AB Dor indicate significant activity, with the FUV flux being the brightest among the stars in our study. The archival GALEX FUV flux of AB Dor is $1.3\times10^{-14}$ ergs s$^{-1}$ cm$^{-2}$ \r{A}$^{-1}$, the highest amongst the sample studied in this paper.

    \item BO Mic: During our $\sim$42 ks exposure, which spans approximately 250 ks, we observed two flares occurring within a few hours, while the remaining light curve showed no major X-ray flux enhancements. During the same time interval, the UV light curve also showed a larger variation in the flux, coinciding with the X-ray events. The X-ray data allowed us to determine the global coronal abundances, temperatures, and emission measures during the different flaring and quiescent states. The observed X-ray luminosity of BO Mic increases from a quiescent value of log~L$_{\mathrm X} \sim 29.83$ to 30.72 during the flare. The quiescent luminosity of BO Mic is consistent with the earlier finding of log~L$_{\mathrm X} \sim 29.94$ \cite{Singh_1999}. Furthermore, \cite{makarov_2003} found  log~L$_{\mathrm X} \sim 31.06$ an order of magnitude higher than our estimate however indicating the excess X-ray might be due to a flaring event similar to our observations.  The quiescent FUV fluxes of BO Mic are $2.1\pm 0.4 \times10^{-15} $ ergs s$^{-1}$ cm$^{-2}$ \r{A}$^{-1}$, consistent with GALEX FUV fluxes of $1.7\times10^{-15}$ ergs s$^{-1}$ cm$^{-2}$ \r{A}$^{-1}$.

   \item DG CVn: We observed two large flare-like events in the SXT light curves; however, poor signal-to-noise ratio prevented us from carrying out detailed temporal-spectral analysis during quiescent and flaring states. The X-ray data allowed us to estimate coronal temperatures of 3 MK and 35 MK, consistent with previous findings during the quiescent state of DG CVn \citet{osten_2007}. Although visual inspection of DG CVn's FUV and NUV light curves indicated 5-10\% modulation in the observed flux with rotation, we did not detect any significant periodicity in our data. The GALEX NUV flux is $1.2\times10^{-15}$ ergs s$^{-1}$ cm$^{-2}$ \r{A}$^{-1}$, consistent with our measured flux.

    \item GJ 3331: The coronal X-ray luminosity measured during our observations is $3.4 \times 10^{29} \mathrm{ergs~s^{-1}}$ consistent with the value reported by \cite{voges_1999}. The best fit model is a two-temperature {\it apec} model with sub-solar abundances for all the elements. The best fit temperatures are 0.44$^{+0.27}_{-0.13}$ keV and 1.0$^{+0.08}_{-0.06}$ keV and the elemental abundances are 0.22 times solar for both the components (See Table 4). The GALEX NUV and FUV are $4.2\times10^{-15}$  and $1.8\times10^{-15}$ergs s$^{-1}$ cm$^{-2}$ \r{A}$^{-1}$, respectively. Our measured FUV and NUV fluxes with UVIT are lower than the GALEX fluxes. GJ 3331 has a close companion (GJ 3332) which is not very well resolved in GALEX since the effective aperture of Galex FUV is about twice of UVIT. Hence the GALEX FUV fluxes are significantly higher than our measured fluxes. Consequently, this star has among the highest values of X-ray luminosity to FUV luminosity ratio, indicating high coronal activity for a relatively slow rotator unless the rotation period is indeed as short as 0.34d instead of 9.8d.
\end{enumerate}

Having investigated the individual stars, we compared their X-ray and FUV fluxes with a set of well-studied, non-planet hosting stars from \cite{france_2018}. These comparison stars encompass a range of spectral types, including F, G, K, and M Dwarfs, and exhibit rotation periods spanning from 0.9 to 100 days. Importantly, both X-ray and UV fluxes have been measured for these stars, making them suitable for our comparative analysis. The X-ray fluxes of these well-studied targets are obtained using ROSAT Position Sensitive Proportional Counters (PSPC) source catalogue \cite{voges_1999}. Using XSPEC version 12.9.1 and WebPIMMS v4.11, we convert the flux from canonical ROSAT energy band 0.1-2.4 keV to 0.3-2.0 keV. 

In Figure \ref{fig:lum} (left panel), the observed X-ray surface flux (0.3-2 keV) of our four targets (filled circles) are compared with the fluxes of the reference targets (open circles and square). To ensure comparability between stars of different sizes, we estimated the surface X-ray fluxes by applying a dilution factor $(\frac{\mathrm{d}}{\mathrm R_{\star}})^2$ to the observed X-ray flux. The radius values needed to calculate the dilution factor were obtained from \cite{france_2018}. Note this plot also includes V405~And, yet another ultrafast rotator and a short period RS~CVn type binary, observed by \emph{AstroSat}. A detailed analysis of this target will be presented in Pathak et al. (in preparation).

In Figure \ref{fig:lum} (right panel), we present the FUV surface flux of BO Mic, DG CVn and GJ 3331 and the comparison stars from \cite{france_2018}. To compare these FUV fluxes with \emph{AstroSat}'s UVIT F148W, we used the Si~IV emission line fluxes ($\lambda$ 1393.75 \r{A}, $\lambda$ 1402.76 \r{A}) observed with HST-STIS  \citet{france_2018} as a proxy. Similar to the X-ray analysis, we calculated surface FUV fluxes (S${\mathrm{FUV}}$) by multiplying the observed FUV flux with the observed bandwidth and the dilution factor of $(\frac{\mathrm{d}}{\mathrm{R{\star}}})^2$.

Overall, the measured surface fluxes at X-ray and FUV wavelengths exhibit a consistent trend, aligning with earlier findings by \cite{stelzer_2013}. Both FUV and X-rays demonstrates a slight trend towards the saturation. Note that there is a scarcity of targets with measured FUV and X-ray fluxes for stars with rotation periods $<1$ day. Continued observations of ultra-fast rotators in both UV and X-ray wavelengths with \emph{AstroSat}, will enable further insight of these trends associated with the observed surface flux of stars with rotation periods $<1$ day. Our current observations provide a foundational basis for conducting comprehensive systematic studies of ultra-fast rotators using multi-wavelength capabilities of \emph{AstroSat}, and from all-sky survey data that would soon become available from \emph {eRosita}.

\begin{landscape}
\begin{table}
\label{tab:Tab3}
\caption{Spectral parameters for the best fit model for the stars observed with the SXT}
\begin{tabular}{lllllllllllll}
\hline
Star Name & Spectral Model & Parameters   &     &     &     &    &    &    &  &  &   \\
        &       & kT$_1$   & Z   &  EM$_1$  & kT$_2$ & EM$_2$ & kT$_3$ & EM$_3$ & $\chi^2_\nu$/dof & Flux & Flux \\
      &    &   keV &   & 10$^{51}$cm$^{-3}$  & keV  & 10$^{51}$cm$^{-3}$   & keV &  10$^{51}$cm$^{-3}$ &      &  (Soft)    &  (Hard) \\
\hline
 {\bf AB~Dor} & tbabs*(apec+apec+apec) & 0.25$^{+0.013}_{-0.011}$  & 0.36$^{+0.10}_{-0.08}$  & 62.1   & 0.93$^{+0.03}_{-0.03}$ & 45.6  & 2.7$^{+1.2}_{-0.5}$ & 22.3  & 1.16/151  & 4.3 & 0.7 \\
  (2016 Jan 15-16) &&&&&&& &\\
 {\bf AB~Dor}  & tbabs*(apec+apec+apec) & 0.27$^{+0.015}_{-0.012}$ & 0.35$^{+0.10}_{-0.08}$ & 64.3   & 0.99$^{+0.03}_{-0.03}$  & 47.2  & 3.6$^{+1.9}_{-0.8}$ & 26.8 &  1.05/156  & 4.5 & 1.0 \\
  (2016 Jan 31-Feb 1) & &   & & & &\\
{\bf BO~Mic}   & tbabs*(apec+apec) & 0.74$^{+0.06}_{-0.17}$ & 0.27$^{+0.14}_{-0.10}$ & 2.1   & 1.46$^{+0.38}_{-0.31}$ & 1.9  & -  & - & 1.04/101  & 0.22 & 0.09 \\
(Quiescence) &\\   
{\bf BO Mic}   & tbabs*(apec+apec) & 0.96$^{+0.10}_{-0.11}$ & 0.50$^{+0.54}_{-0.28}$ & 4.1   & 3.6$^{+1.6}_{-0.8}$ & 15.0  & - & - & 1.21/32  & 1.71 & 1.23 \\
(Flare) &\\
{\bf DG~CVn} & tbabs*(apec+apec) & 0.27$^{+0.04}_{-0.03}$ & 1.0  & 2.71 & 2.83$^{+2.8}_{-1.0}$ & 3.60  & - & - &  1.04/34  & 0.20  & 0.07 \\
{\bf GJ~3331}    & tbabs*(apec+apec) & 0.44$^{+0.27}_{-0.13}$  & 0.22$^{+0.07}_{-0.05}$ & 9.84   & 1.0$^{+0.08}_{-0.06}$ & 31.7  & - &  - & 1.44/55  & 0.76 & 0.07 \\
\hline
\end{tabular}

Notes: All errors quoted are with 90\% confidence. N$_H$ in tbabs is kept fixed at 10$^{20}$ cm$^{-2}$ for all spectral models. Abundance, Z,  is relative to solar values for all the elements in apec models. EM$_1$, EM$_2$ and EM$_3$ are the emission measures corresponding to temperature components kT$_1$, kT$_2$ and kT$_3$ respectively. Fluxes are in units of 10$^{-11}$ ergs cm$^{-2}$ s$^{-1}$ and are quoted for energy bands of 0.3$-$2.0 keV (Soft) and 2$-$6 keV (Hard).

\end{table}
\end{landscape}

\section*{Acknowledgements}
This publication uses the data from the {\it AstroSat} mission of the Indian Space Research Organisation (ISRO). These sources were observed as part of both the SXT guaranteed time program and AO. In particular, the data used were observed from the Soft X-ray Telescope (SXT) developed at TIFR, Mumbai. We thank the SXT and UVIT payload operation centre for verifying, and releasing the data via the ISSDC data archive and providing the necessary software tools. This work has also made use of data from the European Space Agency (ESA) mission {\it Gaia} \footnote{\url{https://www.cosmos.esa.int/gaia}}, processed by the {\it Gaia} Data Processing and Analysis Consortium (DPAC\footnote{\url{https://www.cosmos.esa.int/web/gaia/dpac/consortium}}). Funding for the DPAC has been provided by national institutions, in particular, the institutions participating in the {\it Gaia} Multilateral Agreement.
The work has also made use of software, and/or web tools obtained from NASA's High Energy Astrophysics Science Archive Research Center (HEASARC), a service of the Goddard Space Flight Center and the Smithsonian Astrophysical Observatory.  K. P. Singh thanks the Indian National Science Academy for support under the INSA Senior Scientist Programme. 


\begin{landscape}
\begin{table}
\label{tab:Tab-appendix}
\caption{Best fit spectral parameters for all the models used for arriving at the best model given in Table 4.}
\begin{tabular}{lllllllllllll}
\hline
Star Name & Spectral Model & Parameters   &     &     &     &    &    &    &  &  &   \\
        &       & kT$_1$   & Z   &  EM$_1$  & kT$_2$ & EM$_2$ & kT$_3$ & EM$_3$ & $\chi^2_\nu$/dof & Flux & Flux \\
      &    &   keV &   & 10$^{51}$cm$^{-3}$  & keV  & 10$^{51}$cm$^{-3}$   & keV &  10$^{51}$cm$^{-3}$ &      &  (Soft)    &  (Hard) \\
\hline
{\bf AB~Dor} & tbabs*apec    & 0.85 & 1.0   &  31.2   &       -   &    -   &   &    & 13.9/156 &  & \\
 (2016 Jan 15-16) & tbabs*apec & 0.82  & 0.09 & 162.0  &   -   &   -    &   &    & 2.97/155 &  & \\
    & tbabs*(apec+apec) & 0.75 & 1.0  & 22.0 & 2.35 & 29.0  &  &    &  7.2/154  &  & \\
    & tbabs*(apec+apec) & 0.28$^{+0.02}_{-0.02}$ & 0.16$^{+0.02}_{-0.02}$  & 87.0 & 0.97$^{+0.02}_{-0.03}$ & 97.0  &  &    &  1.72/153  & 4.4 & 0.35\\
& tbabs*(apec+apec+apec) & 0.24$^{+0.01}_{-0.01}$ & 1.0 & 30.0 & 0.93$^{+0.03}_{-0.03}$ & 18.0  & 2.9$^{+0.4}_{-0.3}$ & 24.0   &  1.40/152  & 4.3 & 0.85\\
\hline
 {\bf  AB~Dor} & tbabs*apec    & 0.89 & 1.0   &  32.2   & -   &    -   &   &    & 16.1/161 &  & \\
 (2016 Jan 31-Feb 1) & tbabs*apec & 0.85  & 0.08 & 178.0  &   -   &   -    &  &    & 23.04/160 &  & \\
    & tbabs*(apec+apec) & 0.28 & 1.0  & 36.0 & 1.28 & 33.0  &  &    & 5.2/159  &  & \\
    & tbabs*(apec+apec) & 0.31$^{+0.03}_{-0.02}$ & 0.15$^{+0.02}_{-0.02}$ & 87.0 & 1.04$^{+0.02}_{-0.03}$ & 105  & &  & 1.93/158  & 4.6 & 0.44 \\
    & tbabs*(apec+apec+apec) & 0.25$^{+0.01}_{-0.01}$ & 1.0 & 30.0 & 0.98$^{+0.03}_{-0.03}$ & 17.6  & 3.4$^{+0.5}_{-0.4}$ & 31.0   &  1.28/157  & 5.2 & 1.2\\
    \hline
{\bf BO~Mic} & tbabs*apec    & 0.95 & 1.0   &  1.24   &       -   &    -   &  &    &  2.67/104 &  & \\
 Quiescence & tbabs*apec & 0.93$\pm$0.04  & 0.13$\pm$0.03 & 5.2  &   -   &   -    &  &    & 1.21/103 & 0.37 & 0.04\\
    & tbabs*(apec+apec) & 0.76$\pm$0.04 & 1.0  & 0.8 & 2.1$^{+0.7}_{-0.3}$ & 1.4  &  &    & 1.22/102  & 0.33  & 0.09 \\
    & tbabs*(apec+apec) & 0.74$^{+0.06}_{-0.17}$ & 0.27$^{+0.14}_{-0.10}$ & 2.1   & 1.46$^{+0.38}_{-0.31}$ & 1.9  &  &    &  1.04/101  & 0.22 & 0.09 \\
    \hline
{\bf  BO~Mic} & tbabs*apec    & 1.06 & 1.0   &  6.4   &       -   &    -   &  &    &  10.97/35 &  & \\
 Flare & tbabs*apec & 1.66  & 0.14 & 26.4  &   -   &   -    &  &    & 2.3/34 &  &  \\
    & tbabs*(apec+apec) & 0.92$\pm$0.09 & 1.0  & 2.2 & 3.9$^{+1.1}_{-0.7}$ & 13.7  &  &    & 1.24/33  & 1.66  & 1.39 \\
    & tbabs*(apec+apec) & 0.96$^{+0.10}_{-0.11}$ & 0.50$^{+0.54}_{-0.28}$ & 4.1   & 3.6$^{+1.6}_{-0.8}$ & 15.0  &  &    &  1.21/32  & 1.71 & 1.23 \\
 \hline
 {\bf DG~CVn} & tbabs*apec    & 0.33 & 1.0   &  3.2  &       -   &    -   &  &    &  3.19/36 &  & \\
    & tbabs*apec & 0.74$^{+0.12}_{-0.12}$  & 0.03$^{+0.04}_{-0.02}$ & 15.24  &   -   &   -    &  &    &  1.24/35 &  0.20 &  0.013 \\
    & tbabs*(apec+apec) & 0.27$^{+0.04}_{-0.03}$ & 1.0  & 2.71 & 2.83$^{+2.8}_{-1.0}$ & 3.60  &  &    & 1.04/34  & 0.20  & 0.07 \\
 \hline
 {\bf GJ~3331} & tbabs*apec    & 0.93 & 1.0   &  11.4  &       -   &    -   &  &    &  3.68/58 &  & \\
    & tbabs*apec & 0.91$^{+0.04}_{-0.03}$  & 0.16$^{+0.04}_{-0.03}$ & 43.0  &   -   &   -    &  &    & 1.55/57 &  0.75 &  0.074 \\
    & tbabs*(apec+apec) & 0.84 & 1.0  & 8.3 & 2.4 & 9.8  &  &    & 2.02/56  & 0.66  & 0.16 \\

\hline
\end{tabular}
Note: All errors quoted are with 90\% confidence. N$_H$ in tbabs is kept fixed at 10$^{20}$ cm$^{-2}$ for all spectral models. 
Abundance, Z,  is relative to solar values for all the elements in apec model; 
EM$_1$, EM$_2$ and EM$_3$ are the emission measures corresponding to temperature components kT$_1$, kT$_2$ and kT$_3$ respectively;  Fluxes are in units of 10$^{-11}$ ergs cm$^{-2}$ s$^{-1}$ and are quoted for energy bands of 0.3$-$2.0 keV (Soft) and 2$-$6 keV (Hard).
\end{table}
\end{landscape}

\vspace{-1em}

\bibliographystyle{mnras}
\bibliography{paper}

\end{document}